\documentclass[12pt]{article}
\usepackage{psfig,axodraw}
\def\baselinestretch{1.0}
\advance\textheight by \topskip
\begin{document}
\tolerance=100000
\thispagestyle{empty}
\setcounter{page}{0}
\topmargin -0.1in
\headsep 30pt
\footskip 40pt
\oddsidemargin 12pt
\evensidemargin -16pt
\textheight 8.5in
\textwidth 6.5in
\parindent 20pt
 
\def\baselinestretch{1.5}
\newcommand{\newc}{\newcommand}
\def\preprint{{preprint}}
\def\Ord{\lower .7ex\hbox{$\;\stackrel{\textstyle <}{\sim}\;$}}
\def\OOrd{\lower .7ex\hbox{$\;\stackrel{\textstyle >}{\sim}\;$}}
\def\cO#1{{\cal{O}}\left(#1\right)}
\newc{\order}{{\cal O}}
\def\lag             {{\cal L}}
\def\Lag             {{\cal L}}
\def\lum             {{\cal L}}
\def\R               {{\cal R}}
\def\Rsq             {{\cal R}^{\sq}}
\def\Rst             {{\cal R}^{\st}}
\def\Rsb             {{\cal R}^{\sb}}
\def\M               {{\cal M}}
\def\Oas             {{\cal O}(\alpha_{s})}
\def\Vcal            {{\cal V}}
\def\Wcal            {{\cal W}}
\newc{\be}{\begin{equation}}
\newc{\ee}{\end{equation}}
\newc{\br}{\begin{eqnarray}}
\newc{\er}{\end{eqnarray}}
\newc{\ba}{\begin{array}}
\newc{\ea}{\end{array}}
\newc{\bi}{\begin{itemize}}
\newc{\ei}{\end{itemize}}
\newc{\bn}{\begin{enumerate}}
\newc{\en}{\end{enumerate}}
\newc{\bc}{\begin{center}}
\newc{\ec}{\end{center}}
\newc{\ul}{\underline}
\newc{\ol}{\overline}
\newc{\ra}{\rightarrow}
\newc{\lra}{\longrightarrow}
\newc{\wt}{\widetilde}
\newc{\til}{\tilde}
\def\kr              {^{\dagger}}
\newc{\wh}{\widehat}
\newc{\ti}{\times}
\newc{\Dir}{\kern -6.4pt\Big{/}}
\newc{\Dirin}{\kern -10.4pt\Big{/}\kern 4.4pt}
\newc{\DDir}{\kern -10.6pt\Big{/}}
\newc{\DGir}{\kern -6.0pt\Big{/}}
\newc{\sig}{\sigma}
\newc{\sigmalstop}{\sig_{\lstoppair}}
\newc{\Sig}{\Sigma}  
\newc{\del}{\delta}
\newc{\Del}{\Delta}
\newc{\lam}{\lambda}
\newc{\Lam}{\Lambda}
\newc{\gam}{\gamma}
\newc{\Gam}{\Gamma}
\newc{\eps}{\epsilon}
\newc{\Eps}{\Epsilon}
\newc{\kap}{\kappa}
\newc{\Kap}{\Kappa}
\newc{\modulus}[1]{\left| #1 \right|}
\newc{\eq}[1]{(\ref{eq:#1})}
\newc{\eqs}[2]{(\ref{eq:#1},\ref{eq:#2})}
\newc{\etal}{{\it et al.}\ }
\newc{\ibid}{{\it ibid}.}
\newc{\ibidem}{{\it ibidem}.}
\newc{\eg}{{\it e.g.}\ }
\newc{\ie}{{\it i.e.}\ }
\def \viz{\emph{viz.}}
\def \etc{\emph{etc. }}
\newc{\nonum}{\nonumber}
\newc{\lab}[1]{\label{eq:#1}}
\newc{\dpr}[2]{({#1}\cdot{#2})}
\newc{\lt}{\stackrel{<}}
\newc{\gt}{\stackrel{>}}
\newc{\lsimeq}{\stackrel{<}{\sim}}
\newc{\gsimeq}{\stackrel{>}{\sim}}
\def\lsim{\buildrel{\scriptscriptstyle <}\over{\scriptscriptstyle\sim}}
\def\gsim{\buildrel{\scriptscriptstyle >}\over{\scriptscriptstyle\sim}}
\def\lapp{\mathrel{\rlap{\raise.5ex\hbox{$<$}}
                    {\lower.5ex\hbox{$\sim$}}}}
\def\gapp{\mathrel{\rlap{\raise.5ex\hbox{$>$}}
                    {\lower.5ex\hbox{$\sim$}}}}
\newc{\half}{\frac{1}{2}}
\newcommand {\nnc}        {{\overline{\mathrm N}_{95}}}
\newcommand {\dm}         {\Delta m}
\newcommand {\dM}         {\Delta M}
\def\bra{\langle}
\def\ket{\rangle}
\def\cO#1{{\cal{O}}\left(#1\right)}
\def \DM{{\Delta{m}}}
\newc{\bQ}{\ol{Q}}
\newc{\dota}{\dot{\alpha }}
\newc{\dotb}{\dot{\beta }}
\newc{\dotd}{\dot{\delta }}
\newc{\nindnt}{\noindent}

\newcommand{\medf}[2] {{\footnotesize{\frac{#1}{#2}} }}
\newcommand{\smaf}[2] {{\textstyle \frac{#1}{#2} }}
\def\onesq            {{\textstyle \frac{1}{\sqrt{2}} }}
\def\onehf            {{\textstyle \frac{1}{2} }}
\def\oneth            {{\textstyle \frac{1}{3} }}
\def\twoth            {{\textstyle \frac{2}{3} }}
\def\onefo            {{\textstyle \frac{1}{4} }}
\def\forth            {{\textstyle \frac{4}{3} }}

\newc{\matth}{\mathsurround=0pt}
\def\ML{\ifmmode{{\mathaccent"7E M}_L}
             \else{${\mathaccent"7E M}_L$}\fi}
\def\MR{\ifmmode{{\mathaccent"7E M}_R}
             \else{${\mathaccent"7E M}_R$}\fi}
\newcommand{\s}{\\ \vspace*{-3mm} }

\def \ud { {1 \over 2} }
\def \ut { {1 \over 3} }
\def \td { {3 \over 2} }
\newc{\mr}{\mathrm}
\def\dh {\partial }
\def \cs { cross-section }
\def \css { cross-sections }
\def \cm { centre of mass }
\def \cms { centre of mass energy }
\def \cc { coupling constant }
\def \ccs {coupling constants }
\def \gc {gauge coupling }
\def \gcc {gauge coupling constant }
\def \gccs {gauge coupling constants }
\def \yc {Yukawa coupling }
\def \ycc {Yukawa coupling constant }
\def \pp {{parameter }}
\def \pps {{parameters }} 
\def \ps {parameter space }
\def \pss {parameter spaces }
\def \vv {vice versa }

\newc{\siminf}{\mbox{$_{\sim}$ {\small {\hspace{-1.em}{$<$}}}    }}
\newc{\simsup}{\mbox{$_{\sim}$ {\small {\hspace{-1.em}{$>$}}}    }}



\def \eps {\epsilon}
\newc {\Zboson}{{\mathrm Z}^{0}}
\newc{\thetaw}{\theta_W}
\newc{\mbot}{{m_b}}
\newc{\mtop}{{m_t}}
\newc{\sm}{${\cal {SM}}$}
\newc{\as}{\alpha_s}
\newc{\aem}{\alpha_{em}}
\def \PI{{\pi^{\pm}}}
\newc{\ppbar}{\mbox{$p\ol{p}$}}
\newc{\bbbar}{\mbox{$b\ol{b}$}}
\newc{\ccbar}{\mbox{$c\ol{c}$}}
\newc{\ttbar}{\mbox{$t\ol{t}$}}
\newc{\eebar}{\mbox{$e\ol{e}$}}
\newc{\zzero}{\mbox{$Z^0$}}
\def \gamz{\Gam_Z}
\newc{\wplus}{\mbox{$W^+$}}
\newc{\wminus}{\mbox{$W^-$}}
\newc{\ellp}{\ell^+}
\newc{\ellm}{\ell^-}
\newc{\elp}{\mbox{$e^+$}}
\newc{\elm}{\mbox{$e^-$}}
\newc{\elpm}{\mbox{$e^{\pm}$}}
\newc{\qbar}     {\mbox{$\ol{q}$}}
\def \ewgroup{SU(2)_L \otimes U(1)_Y}
\def \smgroup{SU(3)_C \otimes SU(2)_L \otimes U(1)_Y}
\def \smcolorem{SU(3)_C \otimes U(1)_{em}}

\def \SSM  {Supersymmetric Standard Model}
\def \poincare{Poincare$\acute{e}$}
\def \superspace{\emph{superspace}}
\def \sfs{\emph{superfields}}
\def \superpot{\emph{superpotential}}
\def \csf{\emph{chiral superfield}}
\def \csfs{\emph{chiral superfields}}
\def \vsf{\emph{vector superfield }}
\def \vsfs{\emph{vector superfields}}
\newc{\Ebar}{{\bar E}}
\newc{\Dbar}{{\bar D}}
\newc{\Ubar}{{\bar U}}
\newc{\susy}{{{SUSY}}}
\newc{\msusy}{{{M_{SUSY}}}}


\def\photino{\ifmmode{\mathaccent"7E \gam}\else{$\mathaccent"7E \gam$}\fi}
\def\taugluino{\ifmmode{\tau_{\mathaccent"7E g}}
             \else{$\tau_{\mathaccent"7E g}$}\fi}
\def\mphotino{\ifmmode{m_{\mathaccent"7E \gam}}
             \else{$m_{\mathaccent"7E \gam}$}\fi}
\newc{\gl}   {\mbox{$\wt{g}$}}
\newc{\mgl}  {\mbox{$m_{\gl}$}}
\def \winpm{{\wt W}^{\pm}}
\def \winp{{\wt W}^{+}}
\def \winm{{\wt W}^{-}}
\def \charginopm{{\wt\chi}^{\pm}}
\def \mcharginopm{m_{\charginopm}}
\def \mchpmmin {\mcharginopm^{min}}
\def \chonep {{\wt\chi_1^+}}
\def \chonem {{\wt\chi_1^-}}
\def \chplus {{\wt\chi^+}}
\def \chminus {{\wt\chi^-}}
\def \chonip{{\wt\chi_i}^{+}}
\def \chonim{{\wt\chi_i}^{-}}
\def \chonipm{{\wt\chi_i}^{\pm}}
\def \chonjp{{\wt\chi_j}^{+}}
\def \chonjm{{\wt\chi_j}^{-}}
\def \chonjpm{{\wt\chi_j}^{\pm}}
\def \chonepm{{\wt\chi_1}^{\pm}}
\def \chonemp{{\wt\chi_1}^{\mp}}
\def \mchonepm{m_{\chonepm}}
\def \mchonemp{m_{\chonemp}}
\def \chtwopm{{\wt\chi_2}^{\pm}}
\def \mchtwopm{m_{\chtwopm}}
\newc{\dmchi}{\Delta m_{\wt\chi}}

\def \higgsipm{{\wt H}^{\pm}}

\def \vlsp{\emph{VLSP}}
\def \lspi{\wt\chi_i^0}
\def \mlspi{m_{\lspi}}
\def \lspj{\wt\chi_j^0}
\def \mlspj{m_{\lspj}}
\def \lspone{\wt\chi_1^0}
\def \mlspone{m_{\lspone}}
\def \lsptwo{\wt\chi_2^0}
\def \mlsptwo{m_{\lsptwo}}
\def \lspthree{\wt\chi_3^0}
\def \mlspthree{m_{\lspthree}}
\def \lspfour{\wt\chi_4^0}
\def \mlspfour{m_{\lspfour}}


\def \bino{{\wt B}}
\def \win3{{\wt W_3}}
\def \higgsi0{{\wt H}^{0}}

\newc{\sele}{\wt{\mathrm e}}
\newc{\sell}{\wt{\ell}}
\def \msell{m_{\sell}}
\def \slepone{\wt\ell_1}
\def \mslepone{m_{\slepone}}
\def \smuone{\wt\mu_1}
\def \msmuone{m_{\smuone}}
\def \stauone{\wt\tau_1}
\def \mstauone{m_{\stauone}}
\def \snu{\wt{\nu}}
\def \msnu{m_{\snu}}
\def \msnumu{m_{\snu_{\mu}}}
\def \barsnu{\wt{\bar{\nu}}}
\def \barsnul{\barsnu_{\ell}}
\def \snul{\snu_{\ell}}
\def \mbarsnu{m_{\barsnu}}
\newc{\snue}     {\mbox{$ \wt{\nu_e}                         $}}
\newc{\smu}{\wt{\mu}}
\newc{\stau}{\wt{\tau}}
\newc {\nuL} {\wt{\nu}_L}
\newc {\nuR} {\wt{\nu}_R}
\newc {\snub} {\bar{\wt{\nu}}}
\newc {\eL} {\wt{e}_L}
\newc {\eR} {\wt{e}_R}
\def \slepl{\wt{l}_L}
\def \mslepl{m_{\slepl}}
\def \slepr{\wt{l}_R}
\def \mslepr{m_{\slepr}}
\def \stau{\wt\tau}
\def \mstau{m_{\stau}}
\def \slepton{\wt\ell}
\def \mslepton{m_{\slepton}}
\def \mlhiggs{m_{h^0}}

\def \xr{X_{r}}

\def \sfer{\wt{f}}
\def \msfer{m_{\sfer}}
\def \sq{\wt{q}}
\def \msq{m_{\sq}}
\def \msquleft{m_{\tilde{u_L}}}
\def \msqurht{m_{\tilde{u_R}}}
\def \sql{\wt{q}_L}
\def \msql{m_{\sql}}
\def \sqr{\wt{q}_R}
\def \msqr{m_{\sqr}}
\newc{\msqot}  {\mbox{$m_(\sq_{1,2} )$}}
\newc{\sqbar}    {\mbox{$\bar{\wt{q}}$}}
\newc{\ssb}      {\mbox{$\squark\ol{\squark}$}}
\newc {\qL} {\wt{q}_L}
\newc {\qR} {\wt{q}_R}
\newc {\uL} {\wt{u}_L}
\newc {\uR} {\wt{u}_R}
\def \ul{\wt{u}_L}
\def \mul{m_{\ul}}
\newc {\dL} {\wt{d}_L}
\newc {\dR} {\wt{d}_R}
\newc {\cL} {\wt{c}_L}
\newc {\cR} {\wt{c}_R}
\newc {\sL} {\wt{s}_L}
\newc {\sR} {\wt{s}_R}
\newc {\tL} {\wt{t}_L}
\newc {\tR} {\wt{t}_R}
\newc {\stb} {\ol{\wt{t}}_1}
\newc {\sbot} {\wt{b}_1}
\newc {\msbot} {m_{\sbot}}
\newc {\sbotb} {\ol{\wt{b}}_1}
\newc {\bL} {\wt{b}_L}
\newc {\bR} {\wt{b}_R}
\def \mul{m_{\wt{u}_L}}
\def \mur{m_{\wt{u}_R}}
\def \mdl{m_{\wt{d}_L}}
\def \mdr{m_{\wt{d}_R}}
\def \mcl{m_{\wt{c}_L}}
\def \charml{\wt{c}_L}
\def \mcr{m_{\wt{c}_R}}
\newc{\csquark}  {\mbox{$\wt{c}$}}
\newc{\csquarkl} {\mbox{$\wt{c}_L$}}
\newc{\mcsl}     {\mbox{$m(\csquarkl)$}}
\def \msl{m_{\wt{s}_L}}
\def \msr{m_{\wt{s}_R}}
\def \mbl{m_{\wt{b}_L}}
\def \mbr{m_{\wt{b}_R}}
\def \mtl{m_{\wt{t}_L}}
\def \mtr{m_{\wt{t}_R}}
\def \st{\wt{t}}
\def \mst{m_{\st}}
\newc {\stopl}         {\wt{\mathrm{t}}_{\mathrm L}}
\newc {\stopr}         {\wt{\mathrm{t}}_{\mathrm R}}
\newc {\stoppair}      {\wt{\mathrm{t}}_{1}
\bar{\wt{\mathrm{t}}}_{1}}
\def \lstop{\wt{t}_{1}}
\def \lstopbar{\lstop^*}
\def \hstop{\wt{t}_{2}}
\def \hstopbar{\hstop^*}
\def \mlstop{m_{\lstop}}
\def \mhstop{m_{\hstop}}
\def \lstoppair{\lstop\lstop^*}
\def \hstoppair{\hstop\hstop^*}
\newc{\tsquark}  {\mbox{$\wt{t}$}}
\newc{\ttb}      {\mbox{$\tsquark\ol{\tsquark}$}}
\newc{\ttbone}   {\mbox{$\tsquark_1\ol{\tsquark}_1$}}
\def \tsq {top squark }
\def \tsqs {top squarks }
\def \tsql {ligtest top squark }
\def \tsqh {heaviest top squark }
\newc{\mix}{\theta_{\wt t}}
\newc{\cost}{\cos{\theta_{\wt t}}}
\newc{\sint}{\sin{\theta_{\wt t}}}
\newc{\costloop}{\cos{\theta_{\wt t_{loop}}}}
\def \lsbot{\wt{b}_{1}}
\def \lsbotbar{\lsbot^*}
\def \hsbot{\wt{b}_{2}}
\def \hsbotbar{\hsbot^*}
\def \mlsbot{m_{\lsbot}}
\def \mhsbot{m_{\hsbot}}
\def \lsbotpair{\lsbot\lsbot^*}
\def \hsbotpair{\hsbot\hsbot^*}
\newc{\mixsbot}{\theta_{\wt b}}

\def \mhone{m_{h_1}}
\def \hup{{H_u}}
\def \hdn{{H_d}}
\newc{\tb}{\tan\beta}
\newc{\cb}{\cot\beta}
\newc{\vev}[1]{{\left\langle #1\right\rangle}}

\def \abot{A_{b}}
\def \atop{A_{t}}
\def \atau{A_{\tau}}
\newc{\mhalf}{m_{1/2}}
\newc{\mzero} {\mbox{$m_0$}}
\newc{\azero} {\mbox{$A_0$}}

\newc{\lb}{\lam}
\newc{\lbp}{\lam^{\prime}}
\newc{\lbpp}{\lam^{\prime\prime}}
\newc{\rpv}{{\not \!\! R_p}}
\newc{\rpvm}{{\not  R_p}}
\newc{\rp}{R_{p}}
\newc{\rpmssm}{{RPC MSSM}}
\newc{\rpvmssm}{{RPV MSSM}}


\newc{\sbyb}{S/$\sqrt B$}
\newc{\pelp}{\mbox{$e^+$}}
\newc{\pelm}{\mbox{$e^-$}}
\newc{\pelpm}{\mbox{$e^{\pm}$}}
\newc{\epem}{\mbox{$e^+e^-$}}
\newc{\lplm}{\mbox{$\ell^+\ell^-$}}
\def \branch{\emph{BR}}
\def \branche{\branch(\lstop\ra be^{+}\nu_e \lspone)\ti \branch(\lstop^{*}\ra \bar{b}q\bar{q^{\prime}}\lspone)}
\def \branchmu{\branch(\lstop\ra b\mu^{+}\nu_{\mu} \lspone)\ti \branch(\lstop^{*}\ra \bar{b}q\bar{q^{\prime}}\lspone)}
\def\Ecm{\ifmmode{E_{\mathrm{cm}}}\else{$E_{\mathrm{cm}}$}\fi}
\newc{\rts}{\sqrt{s}}
\newc{\rtshat}{\sqrt{\hat s}}
\newc{\gev}{\,GeV}
\newc{\mev}{~{\rm MeV}}
\newc{\tev}  {\mbox{$\;{\rm TeV}$}}
\newc{\gevc} {\mbox{$\;{\rm GeV}/c$}}
\newc{\gevcc}{\mbox{$\;{\rm GeV}/c^2$}}
\newc{\intlum}{\mbox{${ \int {\cal L} \; dt}$}}
\newc{\call}{{\cal L}}
\def \met  {\mbox{${E\!\!\!\!/_T}$}}
\def \cpv  {\mbox{${CP\!\!\!\!/}$}}
\newc{\ptmiss}{/ \hskip-7pt p_T}
\def \eslash{\not \! E}
\def \etslash{\not \! E_T }
\def \ptslash{\not \! p_T }
\newc{\PT}{\mbox{$p_T$}}
\newc{\ET}{\mbox{$E_T$}}
\newc{\dedx}{\mbox{${\rm d}E/{\rm d}x$}}
\newc{\ifb}{\mbox{${\rm fb}^{-1}$}}
\newc{\ipb}{\mbox{${\rm pb}^{-1}$}}
\newc{\pb}{~{\rm pb}}
\newc{\fb}{~{\rm fb}}
\newc{\ycut}{y_{\mathrm{cut}}}
\newc{\chis}{\mbox{$\chi^{2}$}}
\def \hadron{\emph{hadron}}
\def \nlc{\emph{NLC }}
\def \lhc{\emph{LHC }}
\def \cdf{\emph{CDF }}
\def\dzero{\emptyset}
\def \tevatron{\emph{Tevatron }}
\def \lep{\emph{LEP }}
\def \jets{\emph{jets }}
\def \jet(s){\emph{jet(s) }}

\def\Crs{stroke [] 0 setdash exch hpt sub exch vpt add hpt2 vpt2 neg V currentpoint stroke 
hpt2 neg 0 R hpt2 vpt2 V stroke}
\def\loopdk{\lstop \ra c \lspone}
\def\brloopdk{\branch(\loopdk)}
\def\fourdk{\lstop \ra b \lspone  f \bar f'}
\def\brfourdk{\branch(\fourdk)}
\def\fourdklep{\lstop \ra b \lspone  \ell \nu_{\ell}}
\def\fourdkhad{\lstop \ra b \lspone  q \bar q'}
\def\brfourdklep{\branch(\fourdklep)}
\def\brfourdkhad{\branch(\fourdkhad)}
\def\twodk{\lstop \ra b \chonep}
\def\brtwodk{\branch(\twodk)}
\def\threedkslep{\lstop \ra b \wt{\ell} \nu_{\ell}}
\def\brthreedkslep{\branch(\threedkslep)}
\def\threedksnu{\lstop \ra b \wt{\nu_{\ell}} \ell }
\def\brthreedksnu{\branch(\threedksnu) }
\def\threedklsp{\lstop \ra b W \lspone }
\def\brthreedklsp{\\branch(\threedklsp) }
\def\topdk{t \ra \lstop \lspone}
\def\rpvdk{\lstop \ra e^+ d}
\def\brrpvdk{\branch(\rpvdk)}
\def\fonec{f_{11c}} 
\newc{\mpl}{M_{\rm Pl}}
\newc{\mgut}{M_{GUT}}
\newc{\mw}{M_{W}}
\newc{\mweak}{M_{weak}}
\newc{\mz}{M_{Z}}

\newc{\OPALColl}   {OPAL Collaboration}
\newc{\ALEPHColl}  {ALEPH Collaboration}
\newc{\DELPHIColl} {DELPHI Collaboration}
\newc{\XLColl}     {L3 Collaboration}
\newc{\JADEColl}   {JADE Collaboration}
\newc{\CDFColl}    {CDF Collaboration}
\newc{\DXColl}     {D0 Collaboration}
\newc{\HXColl}     {H1 Collaboration}
\newc{\ZEUSColl}   {ZEUS Collaboration}
\newc{\LEPColl}    {LEP Collaboration}
\newc{\ATLASColl}  {ATLAS Collaboration}
\newc{\CMSColl}    {CMS Collaboration}
\newc{\UAColl}    {UA Collaboration}
\newc{\KAMLANDColl}{KamLAND Collaboration}
\newc{\IMBColl}    {IMB Collaboration}
\newc{\KAMIOColl}  {Kamiokande Collaboration}
\newc{\SKAMIOColl} {Super-Kamiokande Collaboration}
\newc{\SUDANTColl} {Soudan-2 Collaboration}
\newc{\MACROColl}  {MACRO Collaboration}
\newc{\GALLEXColl} {GALLEX Collaboration}
\newc{\GNOColl}    {GNO Collaboration}
\newc{\SAGEColl}  {SAGE Collaboration}
\newc{\SNOColl}  {SNO Collaboration}
\newc{\CHOOZColl}  {CHOOZ Collaboration}
\newc{\PDGColl}  {Particle Data Group Collaboration}

\def\issue(#1,#2,#3){{\bf #1}, #2 (#3) } 
\def\ASTR(#1,#2,#3){Astropart.\ Phys. \issue(#1,#2,#3)}
\def\AJ(#1,#2,#3){Astrophysical.\ Jour. \issue(#1,#2,#3)}
\def\AJS(#1,#2,#3){Astrophys.\ J.\ Suppl. \issue(#1,#2,#3)}
\def\APP(#1,#2,#3){Acta.\ Phys.\ Pol. \issue(#1,#2,#3)}
\def\JCAP(#1,#2,#3){Journal\ XX. \issue(#1,#2,#3)} 
\def\SC(#1,#2,#3){Science \issue(#1,#2,#3)}
\def\PRD(#1,#2,#3){Phys.\ Rev.\ D \issue(#1,#2,#3)}
\def\PR(#1,#2,#3){Phys.\ Rev.\ \issue(#1,#2,#3)} 
\def\PRC(#1,#2,#3){Phys.\ Rev.\ C \issue(#1,#2,#3)}
\def\NPB(#1,#2,#3){Nucl.\ Phys.\ B \issue(#1,#2,#3)}
\def\NPPS(#1,#2,#3){Nucl.\ Phys.\ Proc. \ Suppl \issue(#1,#2,#3)}
\def\NJP(#1,#2,#3){New.\ J.\ Phys. \issue(#1,#2,#3)}
\def\JP(#1,#2,#3){J.\ Phys.\issue(#1,#2,#3)}
\def\PL(#1,#2,#3){Phys.\ Lett. \issue(#1,#2,#3)}
\def\PLB(#1,#2,#3){Phys.\ Lett.\ B  \issue(#1,#2,#3)}
\def\ZP(#1,#2,#3){Z.\ Phys. \issue(#1,#2,#3)}
\def\ZPC(#1,#2,#3){Z.\ Phys.\ C  \issue(#1,#2,#3)}
\def\PREP(#1,#2,#3){Phys.\ Rep. \issue(#1,#2,#3)}
\def\PRL(#1,#2,#3){Phys.\ Rev.\ Lett. \issue(#1,#2,#3)}
\def\MPL(#1,#2,#3){Mod.\ Phys.\ Lett. \issue(#1,#2,#3)}
\def\RMP(#1,#2,#3){Rev.\ Mod.\ Phys. \issue(#1,#2,#3)}
\def\SJNP(#1,#2,#3){Sov.\ J.\ Nucl.\ Phys. \issue(#1,#2,#3)}
\def\CPC(#1,#2,#3){Comp.\ Phys.\ Comm. \issue(#1,#2,#3)}
\def\IJMPA(#1,#2,#3){Int.\ J.\ Mod. \ Phys.\ A \issue(#1,#2,#3)}
\def\MPLA(#1,#2,#3){Mod.\ Phys.\ Lett.\ A \issue(#1,#2,#3)}
\def\PTP(#1,#2,#3){Prog.\ Theor.\ Phys. \issue(#1,#2,#3)}
\def\RMP(#1,#2,#3){Rev.\ Mod.\ Phys. \issue(#1,#2,#3)}
\def\NIMA(#1,#2,#3){Nucl.\ Instrum.\ Methods \ A \issue(#1,#2,#3)}
\def\JHEP(#1,#2,#3){J.\ High\ Energy\ Phys. \issue(#1,#2,#3)}
\def\EPJC(#1,#2,#3){Eur.\ Phys.\ J.\ C \issue(#1,#2,#3)}
\def\RPP (#1,#2,#3){Rept.\ Prog.\ Phys. \issue(#1,#2,#3)}
\def\PPNP(#1,#2,#3){ Prog.\ Part.\ Nucl.\ Phys. \issue(#1,#2,#3)}
\newc{\PRDR}[3]{{Phys. Rev. D} {\bf #1}, Rapid  Communications, #2 (#3)}
\def\PROP(#1,#2,#3){Prog.\ Part.\ Nucl.\ Phys. \issue(#1,#2,#3)}
\def\PS(#1,#2,#3){Phys.\ Scripta \issue(#1,#2,#3)}

\vspace*{\fill}
\vspace{-0.5in}
\begin{flushright}
{\tt JU-PHYSICS/11/06}
\end{flushright}
\begin{center}
{\Large \bf
New signals of a R-parity violating model of neutrino mass at the Tevatron
}\\[1.00
cm]
{\large Amitava Datta$^{}$\footnote{\it adatta@juphys.ernet.in}
{and}
{\large Sujoy Poddar} $^{}$ \footnote{\it
sujoy@juphys.ernet.in
}}\\[0.3 cm]

{\it  Department of Physics, Jadavpur University,
Kolkata- 700 032, India}\\[0.3cm]
\end{center}
\vspace{.2cm}

\begin{abstract}
{\noindent \normalsize}
In a variety of models of neutrino masses and mixings the lighter top squark 
decays into
competing R - parity violating and R - parity conserving channels. Using Pythia we 
have estimated in a model independent way the minimum value of P $\equiv$ 
BR($\lstop \ra c \lspone$) $\times$ BR($\lstop \ra l^+_i b$), where $l_i = 
e^+$ 
and $\mu^+$, corresponding to an observable signal involving the final state
 1$l$ + jets +$\met$ (carried by the neutrinos from the $\lspone$ decay) at Tevatron
Run II. For the kinematical cuts designed in this paper P depends on $\mlstop$ only.
We then compute P for representative choices of the model parameters 
constrained by the 
oscillation data and find that over a significant region of the 
allowed parameter 
space P is indeed larger than $P_{min}$. This signal is complementary to the 
dilepton + dijet signal studied in several earlier experimental and 
phenomenological analyses and may be observed even if BR($\lstop \ra l^+_i b$) 
is an 
order of magnitude smaller than  BR($\lstop \ra c \lspone$). The invariant mass 
distribution of the hardest lepton and the hardest jet may determine $\mlstop$
and reveal the lepton number violating nature of the underlying interaction. The 
invariant mass distribution of the two lowest energy jets may determine $\mlspone$.

\end{abstract}
PACS no: 11.30.Pb, 13.85.-t, 14.60.Pq, 14.80.Ly
\vskip1.0cm
\noindent
\vspace*{\fill}
\newpage

\section{Introduction} \label{intro4} 
~~~Neutrino oscillations in
different experiments have \cite{other} established that the neutrinos
have tiny masses, several orders of magnitude smaller than any other
fermion mass in the Standard Model (~SM~) with massless
neutrinos. Massive Dirac neutrinos can be accommodated in the SM if right
handed neutrinos are introduced as SU(2) singlets. But the corresponding
Yukawa couplings must be unnaturally small. There are several more
aesthetic mechanisms of introducing neutrino masses. We shall briefly
review below two popular approaches based on supersymmetry (SUSY) \cite{susyrev} 
and 
their
possible impacts  on the high priority program of SUSY search at high 
energy accelerators - on the current and future  experiments at Tevatron 
Run II in particular.

 The see-saw mechanism \cite{seesaw} in a grand unified
theory(GUT) \cite{GUT}, with or without supersymmetry(SUSY) \cite{susyrev},
offers a natural explanation of small neutrino masses provided the
neutrinos are Majorana fermions. 
One need not fine tune the Dirac masses to 
unnaturally small magnitudes. Instead a typical neutrino Dirac (Majorana) 
mass in this model is assumed to be  
of the order of the electroweak scale(GUT scale 
($M_G$)). The physical neutrino masses turn out to be proportional to the 
ratio of these scales of widely different magnitudes and are, therefore, 
naturally suppressed.

The observation of neutrinoless double beta decay \cite{double}
will 
provide a strong indirect evidence in favour of the Majorana neutrinos.  
Another hall mark of any GUT is the proton decay \cite{GUT} which has not
been observed so far.  However, all non-supersymmetric GUTs suffer from
the naturalness problem \cite{susyrev} which destabilizes the mass of the
higgs boson essentially due to the same large hierarchy of the two mass scales
responsible for the see-saw.

A supersymmetric GUT (SUSYGUT) cures the hierarchy problem provided the
masses of the sparticles (the supersymmetric partners of the SM particles)  
are $\cal O$ (1 TeV). Thus the exciting program of sparticle searches and
the reconstruction of their masses at the on going ( Tevatron Run II) and
the upcoming (the Large Hadron Collider (LHC) or the International Linear
Collider (ILC)) accelerator experiments have the potential of testing
SUSY. Furthermore, in simple grand desert type models SUSY indeed
facilitates the unification of the three couplings of the SM at a 
scale compatible with the current constraints from proton decay.

It should, however, be emphasized that in the most general framework 
with a chosen GUT group the 
neutrino masses and mixing angles involve  many unknown free parameters 
(e.g, the elements of the  Dirac and Majorana mass matrices). Collider 
experiments 
provides very little information on this sector. On the other hand 
Neutrino data alone cannot fully 
determine these parameters unless additional  assumptions are introduced 
to 
simplify the neutrino mass matrix \cite{altarelli}. Such assumptions 
involve the minimal choice of higgs multiplet in a GUT, imposition of 
additional discrete symmetries etc. Thus the observation of both 
neutrinoless double beta decay and proton decay along with the discovery 
of sparticles with masses at the TeV scale may at best be regarded as a 
circumstantial evidence of an underlying SUSYGUT. There is no simple way 
of relating the measured sparticle masses with the physics of the 
neutrino sector.


Although theoretically the idea of unification of the couplings is rather
appealing there is no compelling experimental evidence in favour of it.  
More importantly if neutrinoless double beta decay is observed but the
proton decay remain illusive, the case for an alternative theory would be
strengthened. Such a theory is provided by the R-parity violating (RPV)  
supersymmetry, where R-parity is a discrete symmetry under which the
particles and the sparticles transform differently \cite{rpv}. It should
be noted that the minimal supersymmetric extension of the SM (MSSM)
naturally contains a R - parity conserving (RPC) as well
as a RPV sector \cite{susyrev}. But the couplings in the latter sector violate both
lepton number and baryon number resulting in catastrophic proton decays.
Thus one option is to impose R-parity as a symmetry and eliminate all RPV
couplings.  This model is generally referred to as the RPC MSSM.

 But there is an alternative. If either baryon number or lepton number
conservation but not both is required by imposing appropriate discrete
symmetries then the catastrophic proton decay is suppressed. Such models
are usually referred to as the RPV MSSM.

The lepton number violating version of the RPV MSSM is more appealing 
since it naturally leads to Majorana masses of the neutrinos 
 \cite{rpv,numassold,subhendu} and neutrinoless double beta decay 
 \cite{rpv}. \footnote {The RPV MSSM can be accommodated in a GUT by
introducing, e.g., non-renormalizable higher dimensional operators;
see the concluding section for further discussions  and references.}

More importantly, the observables in the neutrino sector in this depend 
not only
on the RPV parameters but also on the RPC ones (including the sparticle
masses). Thus the precise determination of the neutrino masses and mixing
angles in neutrino oscillations and related experiments on the one hand
and the measurement of sparticle masses and branching ratios (BRs)
at accelerator experiments on the other, can indeed test
this model quantitatively. In addition the collider signatures of this
model are quite distinct from that of the RPC model (see below). In this
paper our focus will be on a novel signature of a RPV model of $\nu$ mass
which can be probed  at Run II of the Tevatron collider.


In the RPC MSSM the lightest supersymmetric particle (LSP) is necessarily 
weakly interacting, stable and, as a consequence, a carrier of missing 
transverse energy ($\met$). This $\met$ is a hallmark of RPC SUSY. In 
contrast the LSP is necessarily unstable and decays into RPV channels,
violating lepton number in the model under consideration. 
In addition other sparticles can also decay into lepton number violating 
modes providing very novel collider signatures. The BRs of
the latter decays, however, depends on the relative magnitudes of their 
widths and that of the competing RPC channels. Thanks to the LSP decay   
the multiplicity of particles in any event is usually much larger compared 
to the corresponding event in RPC SUSY containing the stable LSP.  
Moreover, since the LSP is not a carrier of $\met$ the 
reconstruction of sparticle masses appears to be less problematic 
 in a RPV model. We shall consider a few examples of such 
reconstructions later.
            
Apparently the stringent constraints from the neutrino data (discussed below)
implies that the RPV couplings must be highly suppressed. This in turn
suggests that the branching ratios (BRs) of the lepton number violating
decays of sparticles other than the LSP will be very small compared to the
competing RPC decays. One notable exception, however, is the direct RPV
decay of the lighter top squark ($\lstop$)
 \cite{nmass,biswarup,naba,shibu}, if it happens to be the next lightest
supersymmetric particle(NLSP) while the lightest neutralino ($\lspone$) is
the LSP. The $\lstop$-NLSP assumption is theoretically
well-motivated due to potentially large mixing effects in the top squark
mass matrix \cite{susyrev}. In this case the RPC decays of $\lstop$ occur only
in higher orders of perturbation theory and are also suppressed (to be
elaborated later). Thus they can naturally compete with the RPV 
decays even if
the latter modes have highly suppressed widths as indicated by the
neutrino data. Thus the competition among different decay modes of the
lighter top squark, which may be the only strongly interacting sparticle
within the striking range of Tevatron Run II, is a hallmark of RPV models
of neutrino mass \cite{shibu}.

No signal of either RPC or RPV has been observed so far. Thus constraints
on the RPV parameters \cite{rpv,allanach} have been obtained from the
experimental data. As in the case of a SUSYGUT, the most general RPV
framework has too many parameters. Thus one usually considers some
benchmark scenarios, each consisting of a minimal set of RPV parameters at
the weak scale \cite{abada,gautam}, which can produce an acceptable
neutrino mass matrix consistent with the oscillation data.

Among the examples in ref. \cite{abada},~we have focused on a specific
model with three trilinear couplings $\lambda'_{i33}$ (~where $i=$ 1,2,3
is the lepton generation index~) which can trigger $\lstop$ decays and 
three bilinear RPV parameters
$\mu_{i}$ at the weak scale.  With the above couplings the neutrino masses
turn out to be proportional to $m_b^2$. A brief review of this model 
 along with notations used here may be found in section 2 of \cite{sujoy}.

The stringent upper bounds \cite{abada,sujoy} on the trilinear (bilinear)  
couplings from neutrino data are $\sim 10^{-4}$ (~$\sim 10^{-4}GeV$~).
Thus almost all collider signatures arising from these couplings except
the LSP decay are expected to be unobservable. As already mentioned, a
notable exception could be the direct RPV decay of the $\lstop$ - NLSP 
via a $\lambda'_{i33}$ 
type coupling \cite{naba,shibu} into a $b$-quark and a charged lepton.

\be
\lstop \ra l_i^+ b, 
\ee
\noindent
where i = e, $\mu$ or $\tau$.

This is so because the competing RPC decay modes in this case  are i) the 
loop induced decay \cite{hikasa}
\be
\lstop \ra c \lspone
\ee
and ii) the four body decay \cite{boehm}
\be
\lstop \ra  b \lspone f \bar f'
\ee
\noindent
where $f - \bar f'$ refers to a pair of light fermions. The last two 
decays occur only in higher orders of perturbation theory and, consequently,
their widths are also highly suppressed. It may be recalled that 
for relatively large values of tan $\beta$ the loop induced decay overwhelms both
the RPV decay \cite{shibu} and the four body decay \cite{boehm,me1,shibu1}.

In the earlier experimental and phenomenological analyses  signals from the 
pair production of $\lstop$-$\lstop^*$
 \cite{biswarup,naba,shibu} followed by RPV decays of both were considered. 
The 
model
independent minimum observable branching ratio (MOBR) of the channel $\lstop
\ra e^+ b$ for Tevatron Run II was also estimated as a function of the lighter
top squark mass ($\mlstop$)  \cite{shibu} by considering the dilepton + 
dijet signal. We shall review these limits and
their consequences in a later section. Moreover, it was also demonstrated that
by reconstructing the two lepton - jet invariant masses the lepton number violating
nature of the decay can be directly established \cite{shibu} for a variety of
$\mlstop$.

In this work we shall concentrate on a new signal which arises if one 
member of the $\lstop$-$\lstop^*$ pair decays via the RPV channel into an 
electron or 
a muon while the other decays via the loop induced channel (Eq.2). The 
second decay is assumed to be followed by the LSP decay via the modes 

\be
 \lspone \ra \nu_i b \bar b ~~ ;~~ i=1,2,3
\ee

The modes in Eq.4 indeed occur with a combined BR of 100\% in all 
models provided the LSP mass
$\mlspone$ is smaller than $m_W$ (the W boson mass) so that its decays
into W, Z or t are kinematically forbidden. In addition it holds to a
high degree of accuracy for any $\mlspone$ if the higgsino components of the 
LSP are highly
suppressed(e.g., if it is almost a pure bino) and it is lighter than 
$m_t$ \footnote {see section 3 for further details.}. Thus the signal 
consists of a very energetic
lepton accompanied by several jets (we do not employ flavor tagging)  
plus a moderate amount of missing energy carried by the neutrinos. We have
generated the signal events by using Pythia(version 6206) \cite{spythia}. 
The background
events, discussed in detail in the next section, are generated either
directly by Pythia or by interfacing Pythia and CalcHEP(version 2.3.7)
 \cite{calchep}. We
then introduce kinematical cuts which optimally suppress the backgrounds.  
Finally we define the model independent product branching ratio (PBR)

\be
P \equiv BR(\lstop \ra l^+ b) \times BR(\lstop^* \ra \bar{c} \lspone),
\ee
\noindent
where $BR(\lstop \ra l^+ b) \equiv BR(\lstop \ra e^+ b) + BR(\lstop \ra \mu^+ 
b)$  and 
$\lspone$ is assumed to decay into a pair of jets and missing energy 
 (Eq.4) with 100 \% BR. 
Our simulations estimate the minimum value of P ($P_{min}$) 
as a function of $\mlstop$ corresponding 
to an observable signal. These estimates are for
an integrated luminosity of 9~  $fb^{-1}$ and $ S/ \sqrt{B}=5 $
where $S$ and $B$ are the total number of signal and background events. We
have used the top squark pair production cross section as given by 
QCD \cite{crosssection} as an input. 

We have also examined the invariant mass of the most energetic lepton and
jet in the signal for various $\mlstop$. The resulting distribution peaks
around the input value of $\mlstop$. This illustrates that the
combinatorial backgrounds are not very damaging. 
The peak, if observed, will
unambiguously demonstrate the lepton number violating nature of the
underlying interaction.

Our estimates also establish that the signals proposed in \cite{shibu} and
that proposed in this paper are complimentary. While former happens to be
the most promising search channel if the RPV BR dominates over that of the
loop induced decay, the latter can potentially reveal the presence of RPV 
interactions 
even if the loop decay overwhelms the RPV decays. These points will be 
further illustrated in the subsequent sections with numerical examples.

We then turn our attention to some specific models. 
We compute $P$ in these  models for the entire RPV 
parameter space allowed by the oscillation data and compare the results 
with $P_{min}$ estimated by our Monte Carlo.  We find that for top squark 
masses within the striking range of the Tevatron, a large region of the 
allowed parameter 
space (APS) yields P greater than the estimated 
$P_{min}$. We have also checked that in significant  regions of this
parameter 
space the BR($\lstop \ra l_i^+ b$) is in fact smaller than the estimated 
MOBR in \cite{shibu} or the updated value given in section 3. 
Thus the signal introduced  in this paper is 
indeed complementary to the ones studied earlier and may turn out to be  
the main discovery channel of $\lstop$. 

The plan of the paper is as follows. In section 2 we present our Monte
Carlo studies of the signal and the backgrounds. This is followed by the
main results of this paper: the model independent estimates of $P_{min}$
as a function of $\mlstop$. A comparison of the viability of this signal
vis a vis that of \cite{shibu} is also included.  In section 3 we compute
the parameter $P$ in some specific models and  revise the  estimated MOBR  
in \cite{shibu} for
representative parameter spaces consistent with neutrino oscillation data.
This study  establishes that the two signals are indeed complementary.
Our conclusions and future outlooks are summarized in the last
section.

\section{The Signal and the SM Background}
\label{numerical}

We have simulated the signal described in the introduction by generating
10$^5$ events using Pythia. The most energetic $l-b$ pair in the final
state comes most of the time from the direct decay of $\lstop$. The
kinematics of this pair is, therefore, independent of $\mlspone$. Using
this feature we have shown that it is possible to choose the optimal
kinematical cuts in such a way that the efficiency of these cuts is
practically independent of $\mlspone$. The size of our signal is completely
determined by i) the cross section of $\lstop$ pair production as given by
QCD \cite{crosssection} and the efficiency which depend only on $\mlstop$
among the SUSY parameters and ii) the model independent input value of P
(Eq.5). We have considered the following backgrounds: \\ \\ ~~$~1.~t \bar
t,~~ 2.~ W^{+} W^{-},~~ 3.~W Z,~~4.~W H,~~ 5.~b \bar b,~~ 6.~t \bar b +
\bar t b,~~ 7.~ W + 2j $\\

Backgrounds 1 to 5 have been simulated by Pythia. Backgrounds 6 and 7 have
been generated by CalcHEP at the parton level. Subsequently initial and final
state radiation, hadronization, decay and jet formation have been implemented
by interfacing with Pythia. All cross sections are calculated by CalcHEP in
the leading order using the CTEQ6M parton density functions \cite{pdf} using
the four flavour scheme. The next to leading order(NLO) corrections would
modify both the signal and the backgrounds by the appropriate K-factors.  For
example, the recent QCD prediction for the K-factor for the $t \bar t$ cross
section, the most dominant background (see Table [1]), is 0.94 to 1.52
depending on the choice of the renormalization scale in the leading order
cross section \cite{ttbar}.  The corresponding number for the signal is
$\approx$ 1.3 \cite{nlo}. Since our main result( the estimated $P_{min}$)  is
based on the ratio $S/\sqrt{B}$, we believe that neglecting the NLO
corrections would not change our conclusions drastically.


For the background from $ W b \bar b$, $~\sigma$ has been computed with
nominal cuts of $P_T > 3 \gev$ and $|\eta|<4.0$ on the parton jets.  This
is to eliminate the soft and collinear processes which are important for
the NLO calculation. Here our main aim is to generate events with high
$p_T$, central jets which can contribute to the background surviving the
cuts listed below.  We have checked that reasonable variations of these
nominal cuts do not influence the final results. 

The backgrounds from $W s \bar s$ and $W c \bar c$ have not been simulated
separately since flavour tagging is not included in our selection criteria
and for our cuts these contributions are expected to be similar to that of
$ W b \bar b$.  Similarly the $ W d \bar d$ and $ W u \bar u$ backgrounds
computed with the same nominal cuts as above are practically identical and
only one of them has been simulated. Unless otherwise mentioned
contributions of $W'$s of both signs are included in Table [1].

We present the kinematical cuts and their efficiencies for the signal and 
the backgrounds in Tables [1] and [2] for $\mlstop$= 180 $\gev$. The cuts
($C_1 - C_3$) are as follows:

\begin{enumerate}

\item  $C_1$:Only events having an isolated lepton ($e $ or $\mu$
of either charge) with $\Delta R(l,j)>0.5$ , $|\eta_l|<2.5$ and $P_T>105 \gev$ 
are accepted.Here $\Delta R(m.n) = \sqrt{\Delta\phi^2 + \Delta\eta^2}$, 
m,n stand for either a lepton(l) or a jet(j).

\item  $C_2$:The number of jets in an accepted event is required to be $n_j > 
2$, where, jets are selected by the toy calorimeter of Pythia 
if $ E_T^{j} > $ 12 $\gev$, $|\eta_j|<2.4$
and  $\Delta R(j_1,j_2)>$ 0.5.

\item  $C_3$:Events with the invariant mass of the two highest $P_T$ 
jets lying  between 100 $\gev$ $< M_{j_1 j_2}<$ 70 $\gev$ are rejected.

\end{enumerate}
 
The cuts are applied in the order shown in the table.  The
efficiencies($\eps_i$s) in Table [1] and Table [2] are defined as
${N_i}^s=\eps_i N^s{_{i-1}}$, where i=1,2,3, ${N_i}^s$ is the number of events
selected after the ith cut out of  ${N_0}^s$ generated events.
The expected number of events $N_i$ in Table [1], where i=1,2,3, is obtained by
multiplying the combined efficiency $\eps( = \eps_1 \eps_2...\eps_i)$ with 
$\sigma~\lum$.
                                                                                
The expected number of signal events S(considering final states with both
$e^{\pm}$ and  $\mu^{\pm}$)
 is given by $4. \eps P \sigma \lum$,
where P is the product BR (see Eq.5) sufficient to yield $S/\sqrt{B}=5$.
In our simulations $\eps$ has been computed by generating 
($\lstop \ra e^+ b$) events only
and we have assumed that the efficiency is the same for electrons and muons.  
Out of the three P's estimated in Table [2] the minimum ($P_{min}$) is
obtained by the combination of $C_1$ and $C_2$ only. This conclusion holds
for other choices of $\mlstop$ (within the kinematic reach of Tevatron Run
II) as well. We have tried a variety of additional selection criteria
including b - tagging not shown in Table [1]. For example, 
we have  tried various  lower  $P_T$ cuts on the hardest jet. But the
corresponding $P_{min}$ turns out to be weaker. On the other hand
a lower $P_T$ cut on second hardest jet yields a  P sensitive 
to the LSP mass and introduces model dependence. We therefore conclude 
that the combination of $C_1$ and $C_2$ is the optimal one.

\noindent

\begin{table}[!htb]
\begin{center}

\begin{tabular}{|c|c|c|c|c|c|c|c|}
       \hline
        Backgrounds &$\sigma (\pb)$& 
$\eps_1 $ &$N_1 $& $\eps_2 $&
$N_2 $ &$\eps_3 $&$N_3 $    \\

        \hline
$t \bar t$ &3.73 &0.0179 &601.0 &0.7791 &468.3 &0.9327 &436.8
\\

        \hline
$ W^{+} W^{-}$ &9.51 &0.00623 &533.2&0.2055 &109.5 &0.7188&78.7
 \\
        \hline
$W Z$ &1.16   &0.00541  &56.5 &0.2683&14.9 &0.6831&10.4\\
        \hline
$W H$ &0.11 &0.0093 &9.2  &0.3211 &3.0&0.6286&1.9\\
        \hline
$b \bar b$ &2.82$\times 10^7$    &0.0 &0.0 & 0.0&0.0 &0.0&0.0\\
        \hline
$t \bar b$ +$\bar t b$ &0.33 &0.00824 &24.47  &0.3398&8.31&0.8 &6.65\\
        \hline
$W b \bar b$ &11.6 &.00126   &131.5 &0.119 &15.7  &0.6667&10.5  \\

        \hline

$W u \bar u$ &73.5 &0.00103   &681.3 &0.2136 &145.5  &0.9545& 138.9 \\
        \hline

\end{tabular}
\end{center}
   \caption{Leading order cross sections of the simulated backgrounds and 
the efficiencies of the cuts 
$C_1$, $C_2$ and $C_3$. }
\end{table}

\begin{table}[!htb]
\begin{center}\

\begin{tabular}{|c|c|c|c|c|c|c|c|}
       \hline
        Signal &$\sigma (\pb)$&
$\eps_1 $ &$P $& $\eps_2 $&
$P $ &$\eps_3 $&$P $    \\

\hline
$\lstop \lstop^*$  &0.41  &0.30473 &0.060 &0.9116 &0.037 &0.7702&0.047\\
        \hline

\end{tabular}
\end{center}
   \caption{The signal cross section for $\mlstop =$ 180 GeV
and efficiencies of the cuts $C_1$, $C_2$ 
 and $C_3$ . We have computed  P ( see Eq.5) after each cut 
from Eq.6 by 
requiring
 $S/\sqrt{B}=5$.}
\end{table}


In  Table [2] P (Eq.5) is calculated 
by the formula:
\begin{eqnarray}
P& = &5 \sqrt {\lum \Sigma \sigma^b \eps^b } \over 4\lum \sigma(\lstop 
\lstopbar) \eps,  
\end{eqnarray}
\noindent
where $\sigma^b$ and  $\eps^b$  denote the cross section 
and  the  combined efficiency (i.e., $\eps_b \equiv N_{selected} /
N_{generated} =\eps_1 \eps_2$)
of the background of type b. 
Similarly  $\eps$ is the combined efficiency for the signal.
The integrated luminosity $\lum$ is taken to be
9 fb$^{-1}$. 

We have also computed the backgrounds due to one valence quark and one sea
quark. Some typical process and the corresponding number of background
events (given in parentheses) subject to the cuts $C_1$ and $C_2$ are:  $
W^{+} d d + W^{-} \bar d \bar d $ (12.4) and $W^{+} u d + W^{-} \bar u
\bar d $ (18.7). The corresponding $\sigma$s have been calculated by
applying nominal cuts of $P_T > 3 \gev$ and $|\eta|<4.5$ on the parton
jets. It can be readily checked that these additional backgrounds hardly 
affect the  estimated $P_{min}$.

We present in Fig.1 $P_{min}$ as a function of $\mlstop$ for  $\mlspone$= 
120 GeV. In Fig.2 the 
variation of $P_{min}$ with $\mlspone$ is shown for $\mlstop$ =180$\gev$.
It follows  that $P_{min}$ is almost insensitive to the LSP mass
as claimed above.


\begin{figure}[!htb]
\vspace*{-4.0cm}
\hspace*{-3.0cm}
\mbox{\psfig{file=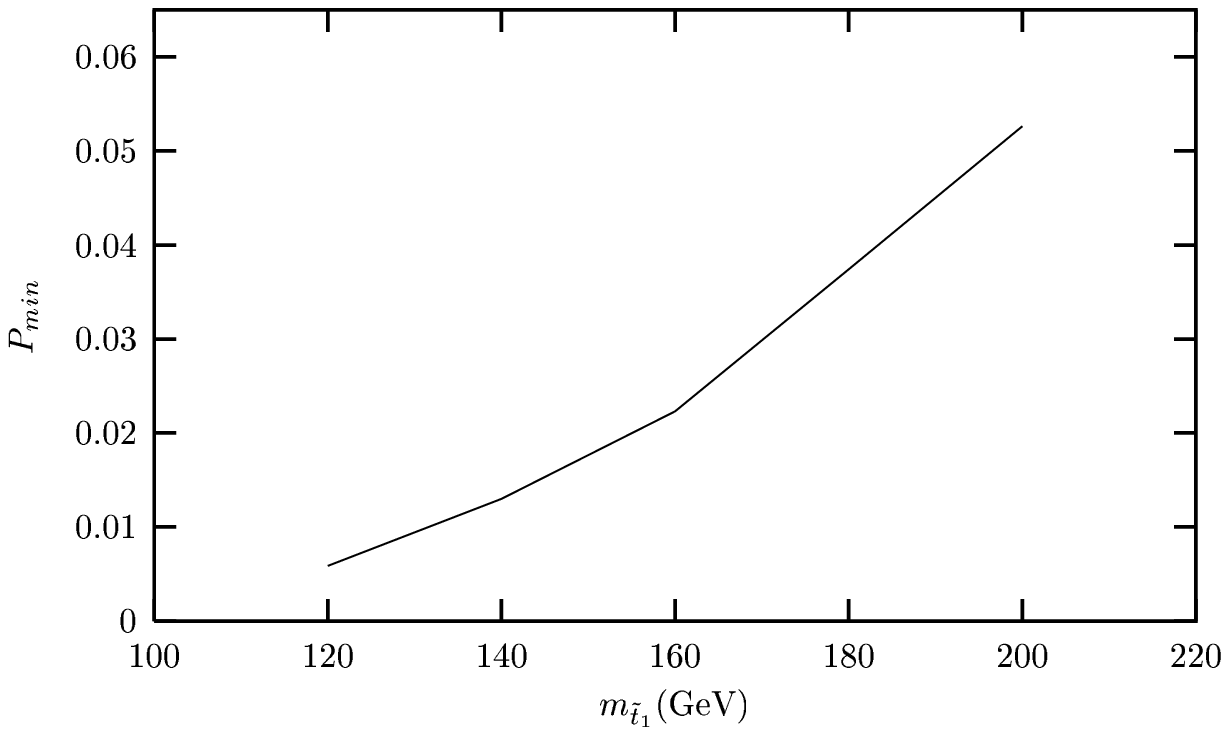,width=20cm}}
\vspace*{-16.2cm}
\caption{The model independent minimum value of the parameter P $\equiv$ 
BR($\lstop \ra c \lspone$) $\times$ BR($\lstop \ra l^+_i b$), where $l^+ = 
e^+$ and 
$\mu^+$, observable at Tevatron Run II via 1$l$ + jets + $\met$ channel 
as a function of $\mlstop$.}
\end{figure}


\begin{figure}[!htb]
\vspace*{-4.0cm}
\hspace*{-3.0cm}
\mbox{\psfig{file=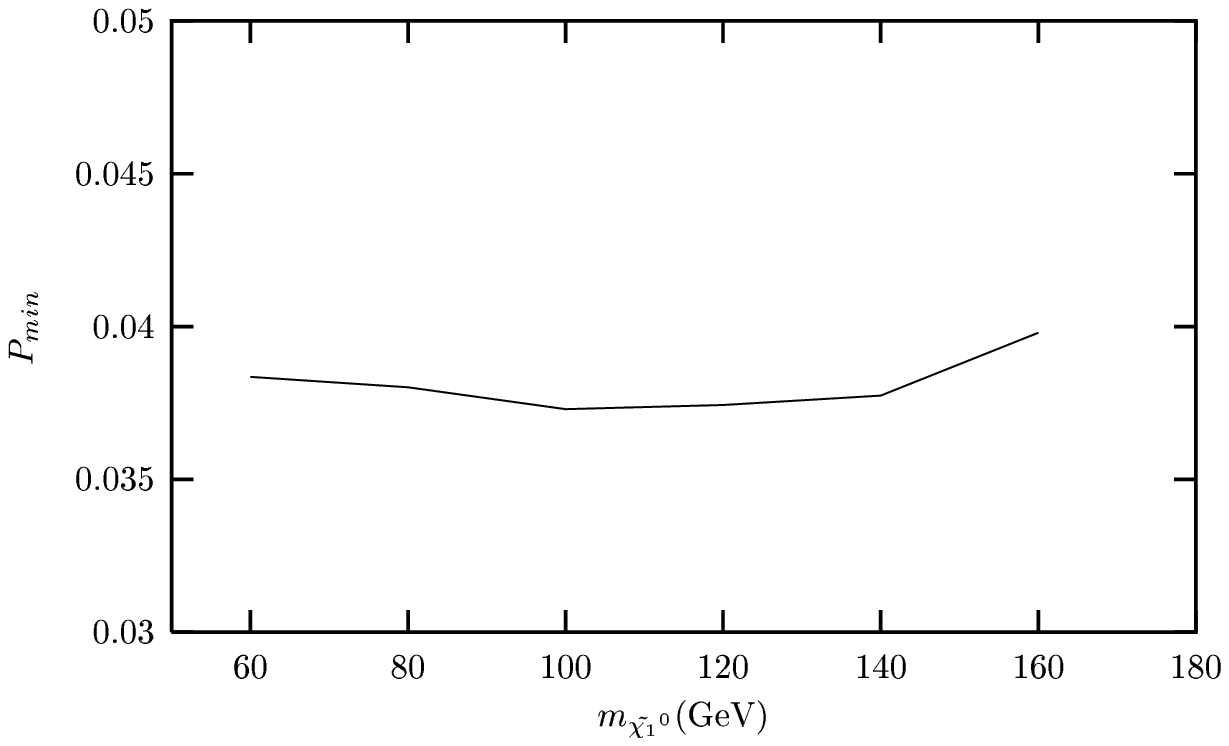,width=20cm}}
\vspace*{-16.2cm}
\caption{The variation of $P_{min}$ (see the caption of Fig.1) with $\mlspone$. }
\end{figure}


\begin{figure}[!htb]
\begin{center}
\hspace*{-1.0cm}   \mbox{\epsfxsize=.5\textwidth
                    \epsffile{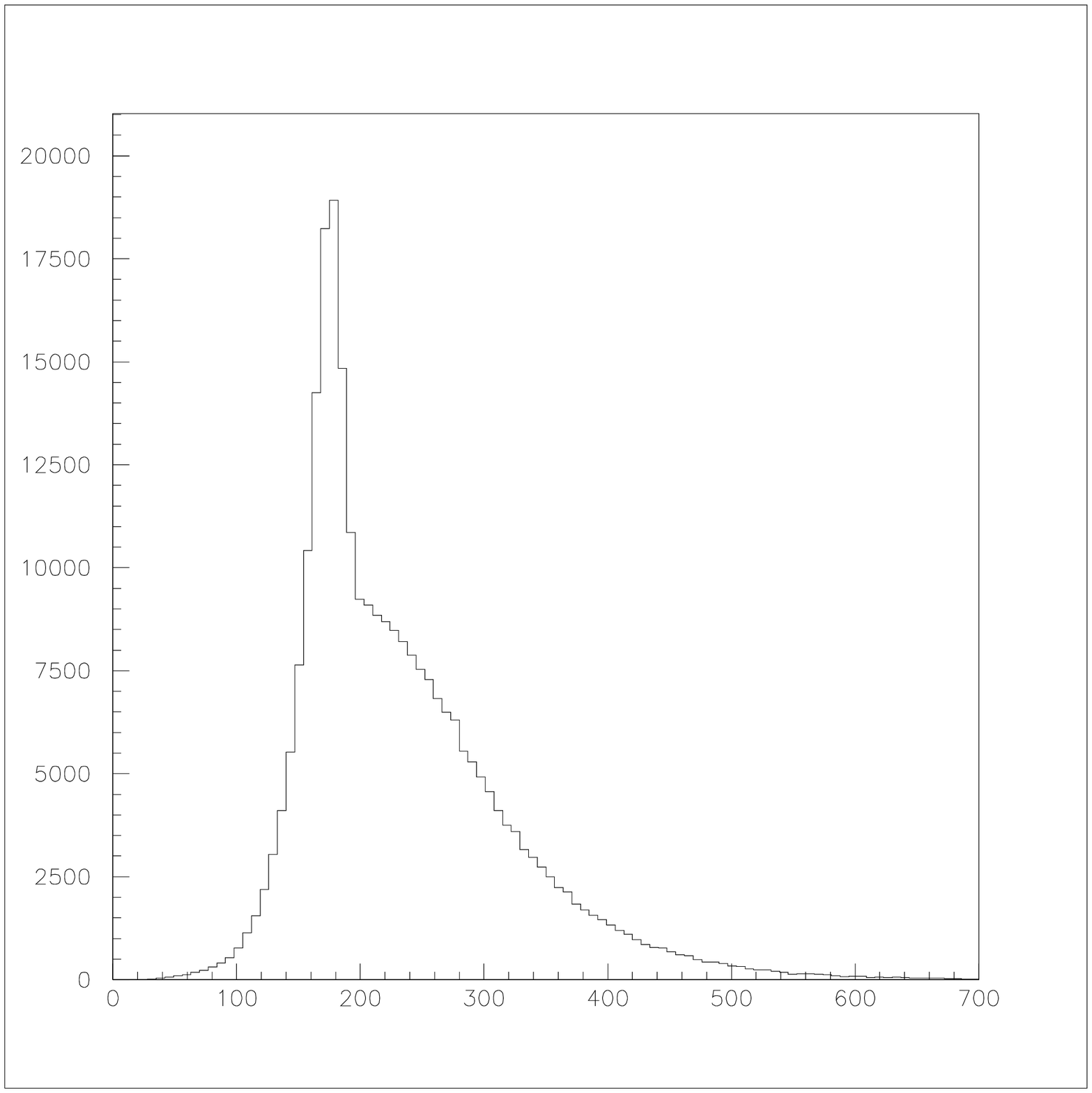}}
\caption{The invariant mass distribution of the hardest jet and the hardest lepton
 in the signal for $\mlstop = 180 \gev$.}
\end{center}
\end{figure}


\begin{figure}[!htb]
\begin{center}
\hspace*{-1.0cm}   \mbox{\epsfxsize=0.55\textwidth
                    \epsffile{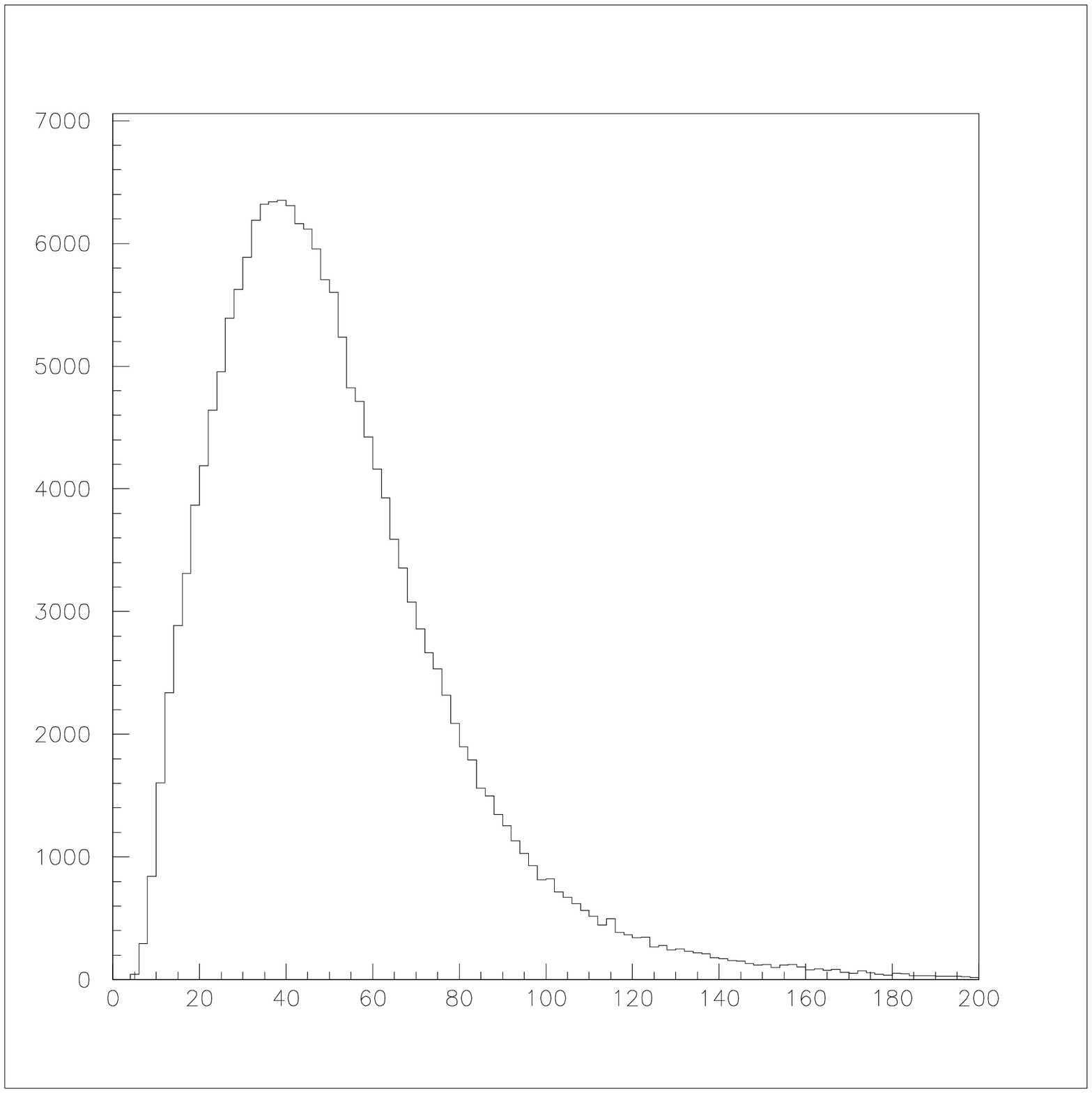}}
\caption{The invariant mass distribution of the lowest two $P_T$ jets in the signal
for $\mlspone \approx 120 \gev$}
\end{center}
\end{figure}


\begin{figure}[!htb]
\vspace*{-4.0cm}
\hspace*{-3.0cm}
\mbox{\psfig{file=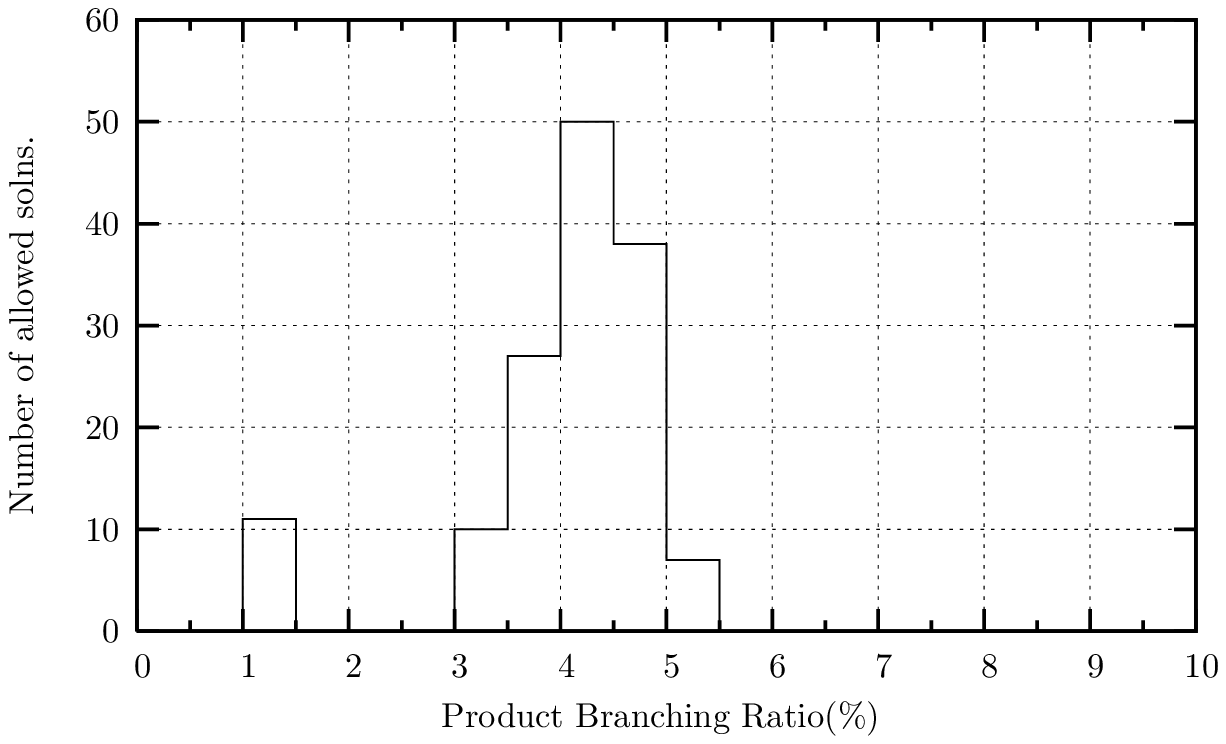,width=20cm}}
\vspace*{-16.2cm}
\caption{ Number of points in the space allowed by the $\nu$ oscillation data vs 
the parameter P($\%$) (Eq.5) in the gaugino like model with $\mlstop$ = 180$\gev$. For the 
choice of parameters and other details see text.  }
\end{figure}


\begin{figure}[!htb]
\vspace*{-4.0cm}
\hspace*{-3.0cm}
\mbox{\psfig{file=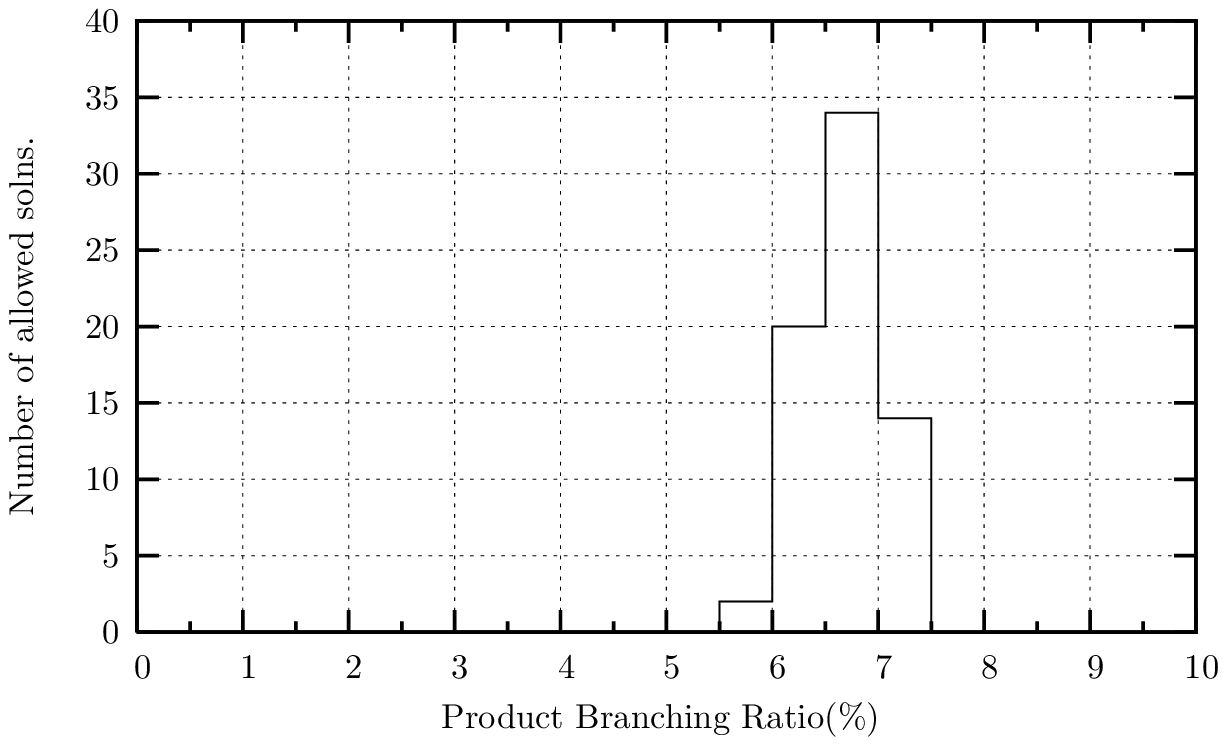,width=20cm}}
\vspace*{-16.2cm}
\caption{ Number of points in the space allowed by the $\nu$ oscillation data vs
the parameter P($\%$) (Eq.5) in the mixed model with $\mlstop$ = 130$\gev$. For the
choice of parameters and other details see text.  }

\end{figure}


We present in Fig.3 the distribution (unnormalized) of
invariant mass of the highest $P_{T}$ lepton
and the highest $P_{T}$ jet in the signal for $\mlstop$ =180$\gev$ and 
$\mlspone$ = 120$\gev$. In spite of the presence of multiple jets in the signal,
the combinatorial background  does not obscure the peak at the input value of
$\mlstop$. In principle $\mlstop$ can be determined from this distribution and
the lepton number violating nature of the underlying
interaction can be established.
Whether this peak will stand over the background depends on the 
actual value of P and $\mlstop$. The issue of the observability of this
peak at Tevatron, therefore, cannot be settled at the moment. 

In Fig.4 we plot the distribution (unnormalized) of 
the invariant mass of the two lowest $P_{T}$ jets in the signal,
which comes most of the time from LSP decay.
The mass of the LSP can be estimated in principle from the upper edge of 
this distribution. Again the combinatorial backgrounds obscure the edge a 
little bit but are not particularly severe.   

Throughout this section we assume BR($\lspone \ra \nu b \bar b$)= 1.0. However, 
if this 
assumption is not strictly true, e.g., due to the presence of several 
rare decay 
modes of the LSP due to  chargino-charge lepton or neutralino 
-neutrino mixing (see the next section) which we have neglected, 
then $P_{min}$ in Fig.1 should 
be interpreted as the minimum value of the product P$\times$ 
BR($\lspone \ra \nu b \bar b$). 

\section{Calculation of the product branching ratio in the model of $m_{\nu}$}

In this section we shall calculate P in some realistic models of neutrino
mass constrained by the neutrino oscillation data 
and examine whether the  predictions  exceed $P_{min}$ 
estimated in the last section. Our main aim is to illustrate that the Run 
II data is sufficiently sensitive to probe these  models and not to make
an exhaustive study of all possible models.

The collider signatures considered in the last section  arise only in 
models with some non vanishing trilinear $\lambda'_{i3j}$ type couplings. 
However, consistency with neutrino oscillation data require the 
introduction of more  RPV parameters (bilinear superpotential terms, 
bilinear soft breaking terms etc). In fact the list of possible choices 
is quite long. It is expected that the 
constraints on the $\lambda'$ couplings in the most general  model 
imposed by the $\nu$ - oscillation data will 
be considerably weaker. On the other 
hand for large $\lambda'$ couplings the dilepton+dijet signal studied
in earlier works \cite{shibu} has a better discovery potential. Thus we 
have 
restricted ourselves to models with a minimal set  of parameters capable 
of explaining the oscillation data with rather stringent constraints on
the $\lambda'$ couplings.

The benchmark scenario studied in this paper was proposed in ref. \cite{abada} 
where it was confronted with the then $\nu$ - oscillation data. We work in a 
basis where the sneutrino vevs are zero. It is assumed that in this basis only 
three nonzero bilinear($\mu_i$) and three trilinear($\lbp_{i33}$) couplings, all 
defined at the weak scale, are numerically significant. In this framework the 
neutrino mass matrix receives contributions both at the tree and one loop level.  
It should be emphasized that the tree level mass matrix yields two massless 
neutrinos. Thus the interplay of the tree level and one loop mass matrices is 
essential for consistency with the oscillation data.

The chargino-charge lepton, the neutralino - 
neutrino and other relevant mixing matrices in this basis may be found in 
\cite{subhendu}. In principle the diagonalization of these matrices may 
induce additional lepton number violating couplings 
which can affect the top squark and LSP decays 
considered in this paper. However, we shall show at the end of this 
section that the decays triggered by such induced couplings are highly 
suppressed either kinematically or dynamically in a wide variety of 
models. As a result 
the approximation that the decays 
of the top squark NLSP  are driven by the $\lbp_{i33}$ 
couplings only is justified. The decay of the LSP requires a more careful
handling and will be taken up at the end of this section.

In \cite{abada} only the upper bounds on $\lbp_{i33}$ couplings 
as obtained from the neutrino oscillation data were 
reported. It was shown 
in refs. 
 \cite{naba}(\cite{shibu}) that the BR limits of the RPV decay of the $\lstop$ 
- NLSP sensitive to Tevatron Run I data ( the MOBR that can be probed by 
Run 
II data) 
correspond to  $\lbp$s which are close to the above upper bounds.  
In ref. \cite{sujoy} the six RPV parameters in 
these models were randomly generated and the neutrino masses and mixing angles 
were 
computed for some well motivated choices of the RPC parameters.
Comparing these with more recent oscillation data \cite{global}a  remarkably 
small allowed parameter space
(APS) was obtained.
It was also shown in \cite{sujoy} that there are six generic 
RPV scenarios 
consistent with the oscillation data \cite{global}. They are : \\

\noindent
a)  $\mu_1 \ll \mu_2, \mu_3 $: \\
($a_1$) $\lbp_{333} >  \lbp_{133}
\geq \lbp_{233}$  \\
($a_2$) $\lbp_{233}> \lbp_{133}  \geq \lbp_{333}$\\

\noindent
b) $\mu_2 \ll \mu_1, \mu_3$: \\
($b_1$) $\lbp_{133} \approx  \lbp_{233} >>  \lbp_{333}$ \\
($b_2$) $\lbp_{233} >  \lbp_{333} >>  \lbp_{133}$ \\

\noindent
c) $\mu_3 \ll \mu_1, \mu_2$:\\
($c_1$) $\lbp_{333} >  \lbp_{133} >>  \lbp_{233}$ \\
($c_2$) $\lbp_{333} >  \lbp_{233} >>  \lbp_{133}$ \\

Each scenario has its characteristic hierarchy among the three leptonic 
BRs of $\lstop$ (see Eq.1). We focus on scenarios $(b_1)$ and $(b_2)$ 
which corresponds to relatively large BR for the decay channel in Eq.1 
with $l_i =$ e or $\mu$. However, in addition to the tree level and $\lbp 
- \lbp$ loop contributions considered in \cite{abada} 
we have included the contribution of the $\mu - 
\lbp$ loops to the $\nu$ - mass matrix. An approximate form of the later 
can be found in\cite{davidson,subhendu}
(see, e.g., Eq.35 of ref.\cite{subhendu}). The inclusion of the new 
contribution does not drastically alter the nature of RPV parameter space 
allowed by the $\nu$-oscillation data.


In addition to the RPV parameters the neutrino masses and mixing 
angles depends on RPC parameters. In this paper we shall
use the following popular assumptions to reduce the number of free 
parameters in the RPC sector: i) At the weak scale the soft breaking mass 
squared parameters of the
L and R-type squarks belonging to the third generation are 
assumed to be the same( the other squark masses are not relevant
for computing neutrino masses and mixing angles in this model).
ii) We shall also use the relation $M_2 \approx 2~ M_1$ at the weak
scale as is the case in models with a unified gaugino mass at $M_G$.
Here $M_1$ and $M_2$ are respectively the soft breaking masses
of the U(1) and SU(2) gauginos respectively. 

The tree level neutrino mass matrix and,hence, the predicted
neutrino masses  depends on the parameters of the gaugino 
sector(through the parameter $C$ \cite{abada,sujoy}). They are $M_2$, $M_1$
, $\mu$ (the higgsino mass parameter) and tan $\beta=v_2/v_1$,
where $v_1$ and $v_2$ are the vacuum expectation values (vevs) for the down 
type and the up type neutral higgs bosons respectively. We remind the 
reader that 
for relatively large tan $\beta$s the loop decay overwhelms the RPV decay.
We have, therefore, restricted ourselves to $tan \beta$ = 5-8. 

It is also
convenient to classify various models of the RPC sector according to the
relative magnitude of $M_2$ and $\mu$. If $M_1 < M_2$ $\ll$ $\mu$, then
the lighter chargino (  $\tilde\chi_1^{\pm}$), the LSP
($\tilde\chi_1^0$) and the second lightest neutralino ($\tilde\chi_2^0$)
are dominantly gauginos. Such  models are referred to as the gaugino-like 
model. On the other hand in the mixed model( $M_1$$ < M_2$$\approx$ $\mu$), 
$\tilde\chi_1^{\pm}$ and $\tilde\chi_2^0$ are admixtures of gauginos
and higgsinos. In both the cases, however, $\lspone$ is almost a bino.
 There are models with $M_1$,$M_2$ $\gg$ $\mu$ 
in which $\tilde\chi_1^{\pm}$,  
$\tilde\chi_1^0$ and $ \tilde\chi_2^0$  are higgsino - 
like  and all have approximately the same mass (
$\approx \mu$).  It is difficult to accommodate the top squark NLSP
in such models  without fine adjustments of the parameters. Thus the LSP 
decay seems to  be the only viable collider signature. 

One can also construct   models  wino or higgsino dominated
LSPs. However, the $\lstop$-NLSP scenario cannot be naturally accommodated in 
these frameworks for reasons similar to the one in  the last paragraph.

The one loop mass matrix, on the other hand, depends on the sbottom
sector (through the parameter $K_2$
 \cite{abada,sujoy}). This parameter decreases for
higher values of the common squark mass for the third generation. From the
structure of the mass matrix it then appears that for fixed C, identical
neutrino masses and mixing angles can be obtained for higher values of the
trilinear couplings if $K_2$ is decreased. Thus at the first sight it
seems that arbitrarily large width of the RPV decays may be accommodated
for any given neutrino data. This, however, is not correct because of the
complicated dependence of the RPV and loop decay BRs of $\lstop$ on the
RPC parameters and certain theoretical constraints. The common squark mass
cannot be increased arbitrarily without violating the top squark NLSP
condition. Of course larger values of the trilinear soft breaking term
$A_t$ may restore the NLSP condition. But larger values of $A_t$ tend to
develop a charge colour breaking( CCB ) minimum of the scalar potential
 \cite{ccb}. Finally the pseudo scalar higgs mass parameter $M_A$ can be 
increased to satisfy the CCB condition. But as noted earlier \cite{sujoy} 
that would enhance the loop decay width as well and suppress the BRs of 
the RPV decay modes.


We first examine scenario$(b_2)$ with the following set of 
RPC parameters:\\ 
A) $M_1=100.0, M_2= 200.0, \mu=320.0, \tan\beta=7.0, A_t=938.0, A_b=300.0,
M_{\tilde q}=400.0, M_{\tilde l}$ (common slepton mass ) = $350.0$
and  $M_A=600.0$,\\  
where all masses and mass 
parameters are in $\gev$ \footnote{For computing the loop decay BR the mass of 
the charm squark is required. We assume it to be equal to the third generation
common squark mass.}. We note in passing that common slepton mass does
not enter into the calculation of the BRs and any choice  which
preserves of $\lstop$ - NLSP condition serves the purpose.
For the
parameters  chosen the loop(4-body) decay BR is $\approx 50 \%(34 \%) $. Yet 
the bulk of the APS gives $P > P_{min}$. 

In this gaugino like model $\mlstop$ = 180$\gev$ and $\mlspone$=100$\gev$.  The 
corresponding  model independent $P_{min}$ is 0.037(see Fig.1).

We then 
randomly generate $10^{7}$ sets of the six RPV parameters under consideration
and count the sets allowed by the oscillation data. For each point of the APS we
compute BRs of the three modes in Eqs.1,2 and 3 and get the corresponding P.  

In Fig.5 we present a histogram of the number of sets  allowed by
the oscillation data vs P in the above  gaugino like model.
It is found that for most of the APS $P> 
P_{min}$. It may be recalled that in ref. \cite{sujoy} it was pointed out that
in this model the loop decay BR is much larger than the total RPV decay BR of 
$\lstop$ even for modest values of $\tan \beta$. Thus the inclusion of the 
$\mu - \lbp$ loop does not change our conclusion drastically.

In Fig.6 we present a similar histogram  in a mixed model. Here the LSP
has a relatively large higgsino component. Thus the additional decay modes
discussed in the introduction (see after Eq.4) may open up and reduce the combined BR of 
the modes in Eq. 4. Hence we have restricted ourselves to $\mlspone <
m_W$.

We have
considered the $(b_1)$ scenario for the RPV parameters.  The RPC parameters
corresponding to this  mixed model are chosen to be:\\ 
B) $M_1=78.0, M_2= 
170.0,
\mu=180.0, \tan\beta=8.0, A_t=890.0, A_b=1000.0,M_{\tilde q}=375.0, 
M_{\tilde
l}=350.0, M_A=300.0,$\\
where all mass and mass parameters are in $\gev$. With the above choice of
parameters $\mlstop$ = 130 GeV and $\mlspone$. Since, as discussed in  
section 2, $P_{min}$ is highly insensitive to $\mlspone$ one can still use
the estimates of figure 2. In this case
the loop(4-body) decay BR is $\approx  85\%(6.5 \%) $. Yet 
the entire APS gives $P > P_{min}$.

We next compare and contrast the signals considered in this paper
and the one discussed in \cite{shibu}. In the latter work assuming both the 
$\lstop$s produced at Tevatron Run II
decay via the $\tilde t_1 \ra e^{+}d$ channel the MOBR of this channel was 
estimated to be 20\%. The Drell-Yan process turned out to be the dominant 
source of background. It was further noted that if b -tagging is employed
the signal will be essentially background free. Using 
the efficiencies as given in Table [2] of ref.\cite{shibu} and  including a 
b-tagging efficiency of 50\% the MOBR  for the channel  $\tilde t_1 \ra 
e^{+}b$ 
corresponding to ten signal events was estimated  to be roughly the same. We 
define the parameter  BR($\tilde t_1 \ra 
e^{+}b$) + BR( $\tilde t_1 \ra \mu^{+}b$)$\equiv$ BR($e + \mu$). Following the
procedure of ref. \cite{shibu} briefly sketched above, we 
then estimate the minimum
observable value of this parameter ($\equiv$ MOBR($e + \mu$)) for $\lum = 9\fb^{-1}$.  
The results are given in Table[3]

\begin{table}[!htb]
\begin{center}\

\begin{tabular}{|c|c|c|c|}
       \hline
        $\mlstop$ &$\sigma (\pb)$&
$\eps$&MOBR($e+\mu$)     \\

\hline
100 & 13.1 & 0.0194 & 0.076 \\
140 & 2.1  & 0.0933 & 0.087 \\
180 & 0.41 & 0.2278 & 0.126 \\
220 & 0.12 & 0.3073 & 0.20  \\  
\hline

\end{tabular} 
\end{center}
   \caption{The minimum observable value of BR($\lstop \ra e^+ b$) +($\lstop \ra
\mu^+ b$) at Tevatron Run II via the dilepton dijet channel for different 
$\mlstop$.}
\end{table}

We find that in any gaugino like model with $\tan \beta \geq 6$ the computed
BR($e +\mu$)s turn out to be smaller than the MOBR in Table [3] practically
over the entire APS . For example, in the gaugino like model considered above
with  $\tan \beta = 7$,  BR($e$) $\approx 0 $ and  BR($\mu$) varies 
from 0.06
to 0.11 (except for a few solutions  which corresponds to 
BR($\mu$) around 0.025). Thus BR($e + \mu$) is indeed much smaller than 
the MOBR in Table [3]  for $\mlstop=180 \gev$. However, most of the points
in the APS yield P larger than $ P_{min}$(Fig.5).  

For the  mixed model considered above the entire APS yields  
P $> P_{min}$. On the contrary BR($e + \mu$) is still 
below the MOBR since the loop and 4-body decay BRs are 
$\approx 85\%$ and $\approx 6.5\%$ respectively. 
These examples illustrate that the signal in
 \cite{shibu} and the one considered in this paper are indeed complimentary. 
 

As noted at the beginning of this section some RPC interactions can induce 
new lepton number violating interactions of $\lstop$ and $\lspone$ 
due to chargino- lepton or 
neutralino-neutrino mixing. These induced interactions can in principle 
affect the decays of the $\lstop$-NLSP or the LSP considered in section 2. 
For example the RPC vertex $\lstop$ - b - $\winpm$ ( or $\higgsipm$) may 
lead to additional lepton number violating couplings of the $\lstop$ due 
to chargino-lepton mixings. Similar induced couplings may arise from 
$\lstop$ - t - $\win3$( or $\bino$ or $\higgsi0$) coupling due to mixing 
in the neutralino-neutrino sector ( these couplings may be relevant only 
if the $\lstop$ is significantly heavier than the t). The mixing 
factors which would suppress the induced couplings can be estimated from 
the 5 $\times$ 5 ( 7 $\times$ 7) chargino - lepton ( neutralino-neutrino) 
mass matrix. The estimated value is $\order(\mu_i / 100 GeV)$ where 100 
GeV is the typical magnitude of an element of the 2 $\times$ 2 chargino 
block or the 4 $\times $4 neutralino block of the above matrices. Since 
the largest $\mu_i$ allowed by the oscillation data is $\order(10^{-4})$ 
GeV the mixing factors are estimated to be $\order(10^{-6})$ or smaller. 
Moreover, the induced couplings will be additionally suppressed by gauge 
or Yukawa couplings. On the other hand the smallest $\lam'_{i33}$ coupling 
contributing to $\lstop$ decay consistent with oscillation data is 
$\order(10^{-5})$. Thus the $\lstop$-NLSP BRs computed by considering 
$\lam'_{i33}$ driven decays only are quite reliable. The rough estimates 
presented here would be substantiated below by results obtained 
by numerically diagonalizing the chargino-lepton and neutralino-neutrino 
mass matrices.

The  LSP decays require more careful analysis. The main decay mode 
considered in this paper (Eq 4) is a three body decay. On the other
hand lepton number violating two body decays of the LSP 
(the decays $\lspone \ra \nu Z $ and $\lspone \ra l^{+} W $ are examples)
can be induced
by the $\lspone$ - $\winp$ - W$^{-}$ or $\lspone$ - $\win3$ - Z vertices.
However, when the LSP is almost a pure bino, which is the case in 
the gaugino model, the original RPC  couplings are highly 
suppressed. In addition 
suppression
by the mixing factors discussed in the last paragraph  will come into 
play.
Consequently the BRs of the lepton number violating 2 body decays of the
LSP are $\order(10.0\%)$ (see below for numerical results). Thus the decay 
in Eq. 4 indeed occur with almost 100 \% BR in this model. 

In more general models with the LSP having significant higgsino components 
(e.g, in the mixed model)
our assumption regarding the LSP decay is valid if $\mlspone < m_W$. For
heavier LSPs the parameter $P_{min}$ in section. 2 should be interpreted as 
the
minimum  value of the product P $\times BR(\lspone \ra \nu_i b
\bar{b})$ observable at Run II.   
  
We now present some numerical results. We numerically diagonalize  the 
mass 
matrices in the chargino-lepton or neutralino-neutrino sector for
all combinations  of RPV parameters allowed by the oscillation data. For 
the parameter set A) (the gaugino like model) we find that  the 
maximum amplitude for 
finding a charge lepton mass eigenstate in a $\winpm$ or $\higgsipm$ is
3.3 $\times 10^{-6}$. The corresponding $\lam'_{233}$, responsible for 
the $\lstop$ or LSP decay signal, is 8.6 $\times 10^{-5}$. Similarly
the maximum amplitude for
finding a neutrino  mass eigenstate in a $\bino$, $\win3$  or $\higgsi0$ is
3.6 $\times 10^{-6}$. The corresponding $\lam'_{233}$ is 8.6 $\times 
10^{-5}$.
Using the induced couplings in this scenario as given above, the 
mixings in the RPC chargino and neutralino mass matrices and the widths of 
the modes ($\lspi \ra Z \lspj $) and ($\lspi \ra W^{\pm} {l_j}^{\mp} $)
given,e.g., in \cite{haber} Eq.15 to Eq.21, we find that
BR($\lspi \ra Z \nu_j $) = 13.0$\%$ and BR($\lspi \ra W^+ l_j $) = 1.0$\%$.

\section{Conclusion}
                                                                                
In conclusion we reiterate that the 1l+ jets($\geq 3$)+$\met$ signal arising 
from
RPC and RPV decays of $\lstop$ - $\lstop^*$ pairs produced at the Tevatron
is a promising channel for probing a class of RPV models of
neutrino mass. For a set of kinematical cuts suitably optimized in this paper 
 the size of this signal is essentially controlled by the
production cross section of the $\lstop$ - $\lstop^*$ pair as given by QCD
and the parameter P (see Eq.5). Using  Monte Carlo simulations 
we have obtained model independent
estimates of $P_{min}$ for different $\mlstop$s corresponding to an observable
signal for an integrated luminosity of 9 $fb^{-1}$ (see Fig.1).  The
efficiencies of the cuts are entirely controlled by the kinematics of the
lepton and the jet with highest $E_T$ which most of the time come directly
from the RPV decay of the $\lstop$. The size of the signal and,
consequently, the estimated $P_{min}$ are  practically
independent of $\mlspone$ (Fig.2).

We have also noted that in spite of the combinatorial backgrounds, the
invariant mass distribution of the above lepton-jet pair shows a peak at
$\mlstop$(see Fig.3). This peak, if discovered, will clearly establish the
lepton number violating nature of the underlying interaction. Similarly
the end point of the invariants mass distribution of the two lowest $E_T$ jets 
(Fig.4) which principally
arise from LSP decays ( Eq.4) may determine
the LSP mass. Whether these salient features of the distributions will 
be obscured by  the full SM background
in the Tevatron data depends on the actual values of P and
$\mlstop$.

This signal may turn out to be the main discovery channel even if the 
loop decay (Eq.2) of the
$\lstop$-NLSP (followed by the LSP decay)
strongly dominates over its RPV decay (Eq.1). On the other hand if the RPV 
decay mode
overwhelms the loop decay then the dilepton + dijet channel studied in
 \cite{biswarup,naba,shibu} may provide a better signal. Finally if
the data establish a competition between  the two modes that would also be 
highly
indicative of an underlying RPV model of neutrino mass. It may be recalled
that in these models the neutrino oscillation data requires the
$\lambda'_{i33}$ couplings to be highly suppressed. As a result the 2-body RPV
decays have widths comparable to the competing RPC decays which occur in 
higher
orders of perturbation theory if $\lstop$ is the NLSP.

The prospect of discovering the RPV model considered in this paper will be
better if the RPV decays of  $\lstop$ into  final states with $\tau +$b can 
also
be probed at the Tevatron. In fact scrutinizing these models in the light of
the oscillation data reveal that $\lstop \ra \tau^+$b is indeed the most 
dominant
decay mode over a large region of the APS \cite{sujoy}. A model independent
estimate of $P_{min}$ for this channel is, therefore, very
important for a complete probe of  this model.
                                                                                
Our computation of P in specific benchmark models \cite{abada,gautam} of
neutrino oscillations  establish that the APS filtered out by the
oscillation data contain  many points with P$> P_{min}$. This happens 
both in the gaugino like( Fig.5) and in the mixed model (Fig.6). There are 
also
regions in the APS where the dilepton + dijet signal 
proposed in \cite{shibu} is unobservable 
but the signal proposed in this paper stands over the background. Thus the two
signals are indeed complementary.

Since in the RPV MSSM leptons are baryons must be treated differently there is a
basic incompatibility between this model and a typical GUT in which quarks and
leptons are placed in the same multiplet. Thus RPV terms in the superpotential
tends to violate both baryon and lepton number conservation and lead to
catastrophic proton decays. Thus the task of a model builder is to remove
such terms by introducing appropriate discrete
symmetries \cite{hall}.  
Yet the RPV MSSM can be accommodated in the framework of a GUT.
RPV interactions may, for example, be induced at the GUT scale by higher
dimensional non-renormalizable operators which reduce to either baryon number 
or lepton number violating interactions when the GUT breaks down to the SM and
certain heavy scalar fields develop vevs \cite{brahm}.    

In GUT models neutrino masses 
can be generated at the weak scale in a variety of interesting ways. The
set of input RPV parameters at $M_G$ need not be identical to the set
appearing in the neutrino mass matrix at the weak scale. In fact the former
set may have smaller number of parameters than the latter set. For example, one may start 
at $M_G$ with
three relatively large trilinear couplings different from the 
$\lambda'_{i33}$s required by the neutrino sector.
Renormalization group evolution \cite{rgevol} and flavour violation inevitably
present in any model due to the CKM mixing would then induce the
$\lambda'_{i33}$ and the $\mu_{i}$ parameters at the weak scale
 \cite{anirban}. However, other RPV parameters may also be generated leading 
to a more complicated $\nu$- mass matrix.
 The relatively large input couplings may then lead to a rich
low energy phenomenology \cite{anirban} in addition to
$\lstop$ and LSP decays. In this paper, however, we have not considered the
origin of the weak scale parameters and have restricted ourselves to the
signatures of the $\lambda'_{i33}$ couplings only.

The signal discussed in this paper and the one in ref. \cite{shibu} will
certainly have much larger sizes at the LHC. But at higher energies many other
sparticles may be produced as well. Thus one has to isolate the signal not
only from the SM background but also from the SUSY background. On the
other hand since the $\lstop$ - NLSP may very well be the only strongly
interacting sparticle within the kinematic reach of the Tevatron. As a result
 these signals may be observed in a relatively clean environment.

\vspace*{5mm}
{\bf Acknowledgments}: \\

SP thanks Dr. S. P. Das for computational help
and Council of Scientific and Industrial Research (CSIR), India for a research 
fellowship. AD acknowledges support from the Department of Science and Technology 
(DST), India under the Project No (SR/S2/HEP-18/2003). 


\end{document}


